\documentclass{PoS}

\usepackage{booktabs}
\usepackage{mathrsfs,latexsym}

\newcommand{\eq}[1]{Eq.~(\ref{#1})}
\newcommand{\fig}[1]{Fig.~\ref{#1}}
\newcommand{\tab}[1]{Table~\ref{#1}}

\newcommand{\tmtextbf}[1]{{\bfseries{#1}}}

\newcommand{\tmtexttt}[1]{{\ttfamily{#1}}}

\newcommand{\Nf}{N_{\mathrm{f}}}

\newcommand{\fm}{\,\mathrm{fm}}

\newcommand{\Mc}{M_{\rm c}}

\newcommand{\Lminus}{{\mathscr L}^{ (\Nf-1) }_{\rm QCD}}
\newcommand{\Leffminus}{{\mathscr L}^{ (\Nf-1) }_{\rm eff}}
\newcommand{\Leffmminus}{{\mathscr L}^{ (\Nf-2) }_{\rm eff}}

\newcommand{\Lsix}{{\mathscr L}_{6}}
\newcommand{\bg}{b_{\rm g}}
\newcommand{\mbarsf}{\kern1pt\overline{\kern-1pt m\kern-1pt}\kern1pt_{{\rm SF}}}
\newcommand{\msbar}{{\rm \overline{MS\kern-0.05em}\kern0.05em}}
\newcommand{\mps}{m_{\rm PS}}
\newcommand{\za}{Z_{\rm A}}
\newcommand{\zp}{Z_{\rm P}}
\newcommand{\ev}[1]{\left\langle #1 \right\rangle}
\newcommand{\geff}{g_{\mathrm{eff}}}
\newcommand{\meff}{m_{\mathrm{light,\,eff}}}

\title{Physical and cut-off effects of heavy sea quarks} 

\ShortTitle{Effects of heavy charm-like quarks}

\author{\speaker{Francesco Knechtli}, Jacob Finkenrath, Bj{\"o}rn Leder\\
        Department of Physics, Bergische Universit{\"a}t Wuppertal\\
        Gaussstr.~20, 42119 Wuppertal, Germany\\
        E-mail: \email{knechtli@physik.uni-wuppertal.de},
                \email{finkenrath@physik.uni-wuppertal.de},
                \email{leder@physik.uni-wuppertal.de}}

\author{Andreas Athenodorou\\
        University of Cyprus, Department of Physics\\
        P.O. Box 20537, Nicosia CY, Cyprus\\ 
        E-mail: \email{athenodorou.andreas@ucy.ac.cy}}

\author{Mattia Bruno, Rainer Sommer\\
        John von Neumann Institute for Computing (NIC)\\
        DESY, Platanenallee~6, 15738 Zeuthen, Germany\\
        E-mail: \email{Mattia.Bruno@desy.de}, \email{rainer.sommer@desy.de}}

\author{Marina Marinkovic\\
        CERN, Physics Department, 1211 Geneva 23, Switzerland\\
        E-mail: \email{marina.marinkovic@cern.ch}}

\abstract{We simulate a theory with two dynamical O($a$) improved Wilson quarks whose mass $M$ ranges from a factor eight up to a factor two below the charm quark mass and at three values of the lattice spacing ranging from 0.066 to 0.034 fm. This theory is a prototype to study the decoupling of heavy quarks. We measure the mass and cut-off dependence of ratios of gluonic observables defined from the Wilson flow or the static potential. The size of the 1/$M$ corrections can be determined and disentangled from the lattice artifacts. The difference with the pure gauge theory is at the percent level when {\em two} quarks with a mass of the charm quark are present.
\begin{flushright} WUP 14-14\\
      DESY 14-209\\
      SFB/CPP-14-86\\
      CERN-PH-TH-2014-215 \end{flushright}
}

\FullConference{The 32nd International Symposium on Lattice Field Theory,\\
		23-28 June, 2014\\
		Columbia University New York, NY}

\begin{document}

\section{Decoupling of heavy quarks}

Heavy quarks are expected to decouple from low energy observables.
We consider here only observables where the quarks with mass $M$
contribute through loops
and no states with an explicit heavy quark.
At energies $E$ much smaller than the mass $M$ of the heavy quark, 
the theory with $\Nf$ quarks, of which one is heavy, is described
by an effective Lagrangian with $\Nf-1$ quark fields
\begin{equation}\label{e:Lminus}
         \Leffminus = \Lminus(\psi_\mathrm{light},\bar\psi_\mathrm{light},A_\mu; 
         \geff(M),\meff(M)) + {1\over M^2}\Lsix \,.
    \end{equation}
In this sense we denote by $\Nf \to \Nf - 1$ the decoupling of one heavy quark.
Here we assume a situation where in finite volume there is a non-anomalous
chiral symmetry in the sector of the light quarks
(which is spontaneously broken in infinite volume).
The terms in the effective Lagrangian have to be
gauge-, Euclidean- and chiral-invariant.
In particular chiral symmetry forbids a dimension five Pauli term.\footnote{
This term was included in the talk in \eq{e:Lminus}.
Its absence was pointed out to us by Martin L\"uscher, whom we thank.} 
$\Lsix$ contains composite fields of dimension six.
The Pauli term only appears in $\Lsix$ in the combination
$m_\mathrm{light} \bar \psi_\mathrm{light} \sigma_{\mu\nu} F_{\mu\nu} \psi_\mathrm{light}$
when the light quarks have a mass $m_\mathrm{light}$.

At low energies the decoupling of the heavy quark leaves traces through renormalization.
The gauge coupling $\geff(M)$ of the effective theory depends
on the mass $M$ of the heavy quark. This dependence comes from the
matching of the effective theory with the full theory. When we assume
all light quarks to be mass-less and neglect
all terms $O(E^2/M^2)$, $\geff$ is the only coupling in \eq{e:Lminus}. 
Only it has to be matched.
However, the value of the coupling at a
given scale is equivalent to the $\Lambda$-parameter of the theory.
This drops out \cite{Bruno:2014ufa}
in all dimensionless physical quantities such as ratios
\begin{equation}
 R(M) = {t_0(M) \over w_0^2(M)} \,,\; {r_1(M) \over r_0(M)}\,,\; \ldots
\end{equation}
Thus for such ratios, the matching is irrelevant.
In the same way, the value of the bare improved coupling\footnote{
By $\tilde{g}_0^2=(1+\bg(g_0^2)\,{am})\,g_0^2$ we denote the improved bare 
gauge coupling of the fundamental theory formulated on the lattice, where 
$m$ is the bare PCAC mass and $g_0$ is the bare gauge coupling.}
$\tilde{g}_0^2$
is irrelevant when one computes ratios such as $r_1/r_0$ with $r_1/a$ and
$r_0/a$ at the same mass.
Therefore these ratios are directly given by their values in the theory with the
heavy quark removed up to power corrections in $1/M^2$. It is the size
of these corrections which we want to estimate here.
In the following we denote by $M$ the renormalization group invariant
quark mass \cite{Capitani:1998mq}.
\begin{table}[tph]
 \centering
\begin{tabular}{cccccc}
\toprule 
$\beta$ & $a$ [$\fm$]  & BC & $T\times L^3$   & $M/\Lambda_\msbar$ & kMDU\\
\midrule
$5.3$ & $0.0658(10)$ & p &$64\times 32^3$   & 0.638(46) & 1 \\
      &              & p &$64\times 32^3$   & 1.308(95) & 2 \\
      &              & p &$64\times 32^3$   & 2.60(19)  & 2 \\
\midrule
$5.5$ & $0.0486($\phantom{0}$7)$  & o & $120\times 32^3$& 0.630(46) & 8 \\ 
      &              & o & $120\times 32^3$& 1.282(93) & 8\\ 
      &              & p & $96\times 48^3$ & 2.45(18) & 4 \\
\midrule
$5.7$ & $0.0341($\phantom{0}$5)$  & o & $192\times 48^3$& 1.277(94) & 4 \\
      &              & o & $192\times 48^3$& 2.50(18) & 8 \\
\bottomrule
\end{tabular}
 \caption{Overview of the ensembles. The entries in the column BC refer to
periodic (p) or open (o) boundary conditions.
The values of $M/\Lambda=0.63$, $1.28$ and $2.50$
correspond approximately to $M=200$, $400$ and $800$ MeV.
}
 \label{t:ens}
\end{table}

\section{Simulations}

\begin{figure}\centering
  \resizebox{7cm}{!}{\includegraphics{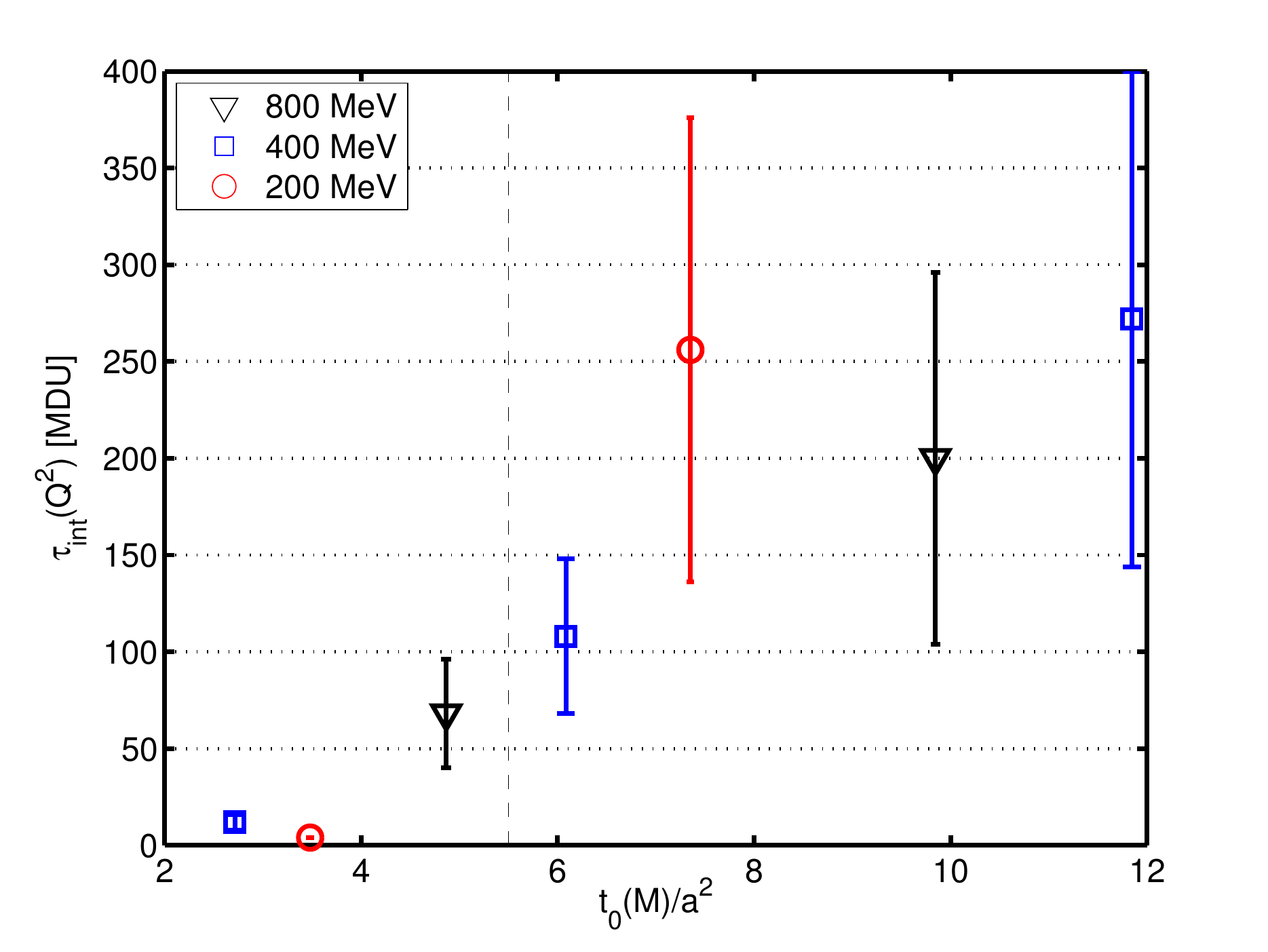}} \ \
  \resizebox{7cm}{!}{\includegraphics{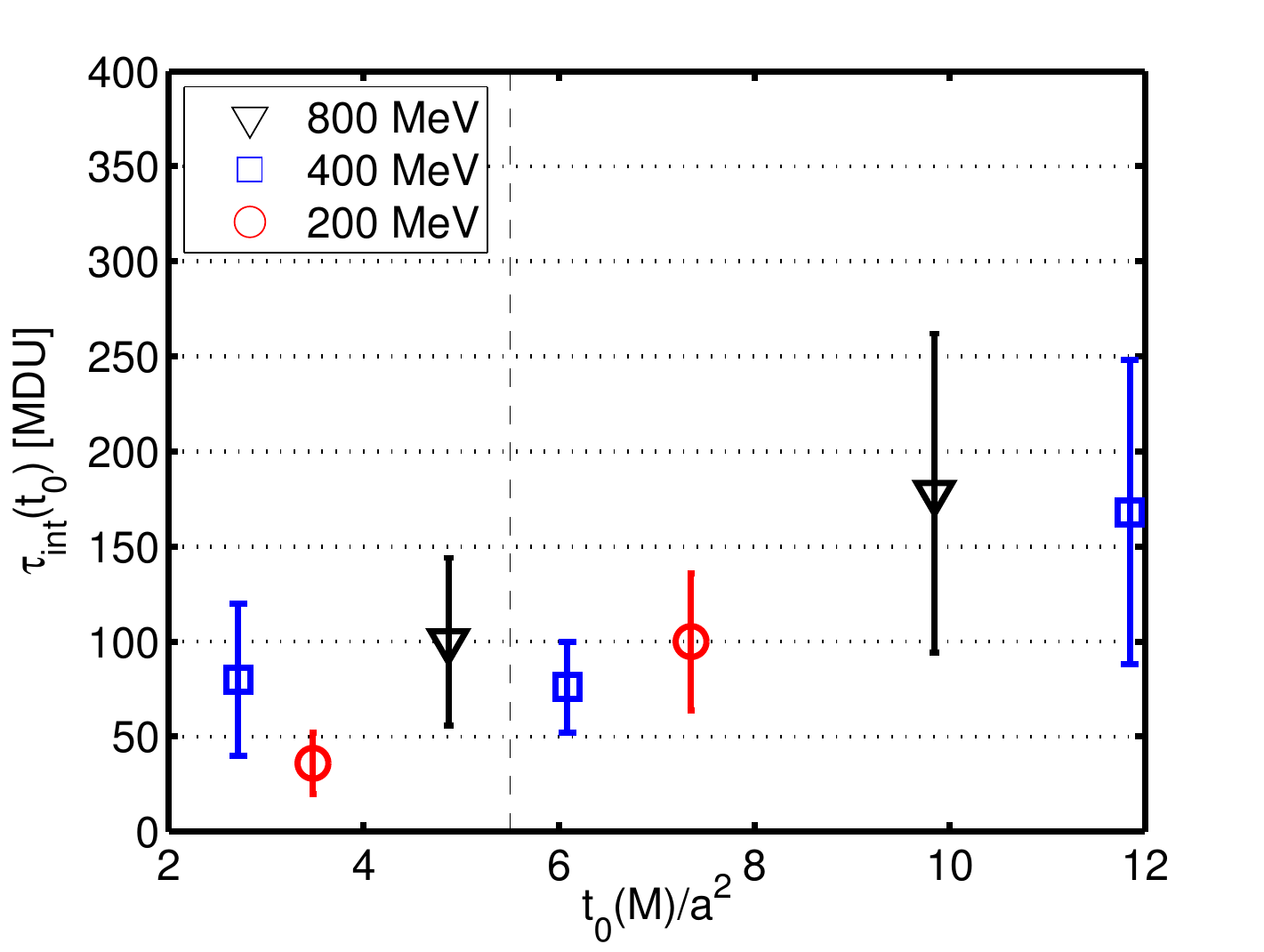}}
  \caption{Autocorrelation times of the topological charge
squared $Q^2$ (left) and of the scale $t_0(M)$ (right) as a function of
$t_0(M)/a^2$. Simulations with $t_0(M)/a^2>5.5$ have been performed with
open boundary conditions.}
  \label{f:taui}
\end{figure}

We simulate a model,
namely QCD with two heavy, mass-degenerate quarks. The decoupling is
then $2\to0$ and the effective theory,
$\Leffmminus$, is the Yang-Mills
theory up to $1/M^2$ corrections.
We use $\Nf=2$ O($a$) improved Wilson fermions \cite{impr:csw_nf2} 
at three values of the
lattice coupling $g_0^2$ and lattice spacing $a$: 
$6/g_0^2=5.3$ ($a=0.0658(10)\,\fm$ from \cite{alpha:lambdanf2}), 
$6/g_0^2=5.5$ ($a=0.0486(7)\,\fm$ from \cite{alpha:lambdanf2}) and
$6/g_0^2=5.7$ ($a=0.0341(5)\,\fm$ estimated from the ratio of the 
scale $t_0/a^2$ at $6/g_0^2=5.5$ and $5.7$).
Our volumes are such that the lightest pseudo-scalar mass times the box size is
$\mps L \ge 7.4$ and $L/r_0(M) \ge 3.8$,
where significant finite volume effects can be excluded.
A list of the simulated ensembles is given in \tab{t:ens}.

Part of the simulations are performed using periodic boundary conditions (except for anti-periodic boundary conditions in temporal direction for the fermions) 
and the MP-HMC algorithm \cite{Marinkovic:2010eg}.
In order to avoid the freezing of the topological charge, for
simulations with $t_0/a^2>5.5$ \cite{algo:csd,Bruno:2014ova} we adopt open boundary conditions
in time and
use the publicly available openQCD package \cite{algo:openQCD}.
At the smallest lattice spacing $a=0.034\,\fm$ we find autocorrelation times
for observables such as $t_0$ or the topological charge squared of
$\tau_{\rm exp}\simeq250$ MDU (Molecular Dynamics Units), see
\fig{f:taui}.
Our statistics of $4000$--$8000$ MDU is therefore adequate.
The error analysis, based on \cite{Wolff:2003sm},
nevertheless includes the effects of modes with 
these large autocorrelation times \cite{algo:csd}.
The cost of our simulations is relatively low compared to
simulations in the chiral regime.

The renormalized quark mass $\mbarsf(L_1)$ at length scale $L_1$ is defined 
by 
$\mbarsf(L_1)=\za/\zp(L_1)\, m$, where the renormalisation
factor $\zp(L_1)$ is defined in the Schr\"odinger Functional scheme
as in \cite{alpha:lambdanf2}. The axial current renormalization factor, $\za$,
is fixed by a chiral Ward identity \cite{DellaMorte:2008xb}. For the determination of the PCAC masses
we use Tomasz Korzec's program\footnote{
It is available at {\tt https://github.com/to-ko/mesons}.}.
The renormalization group invariant mass $M$ is obtained by
multiplying $\mbarsf(L_1)$ with the factor \cite{mbar:nf2} $M/\mbarsf(L_1)=1.308(16)$.
The ratio
\begin{equation}
\frac{M}{\Lambda} = \frac{(aM)\,(L_1/a)}{(\Lambda L_1)}
\end{equation}
is computed using the values of $L_1/a$ from~\cite{alpha:lambdanf2}
(at $6/g_0^2=5.7$ we get $L_1/a=11.07(17)$) and $\Lambda\,L_1$ from
\cite{alpha:nf2}. We take the $\Lambda$ parameter to be defined in the 
$\msbar$ scheme.
The values of $M/\Lambda$ are tabulated in \tab{t:ens}.
Their accuracy is around 7\% and is dominated
by the relative error of $\Lambda\,L_1$.

\section{Numerical results}

\begin{figure}\centering
  \resizebox{7cm}{!}{\includegraphics{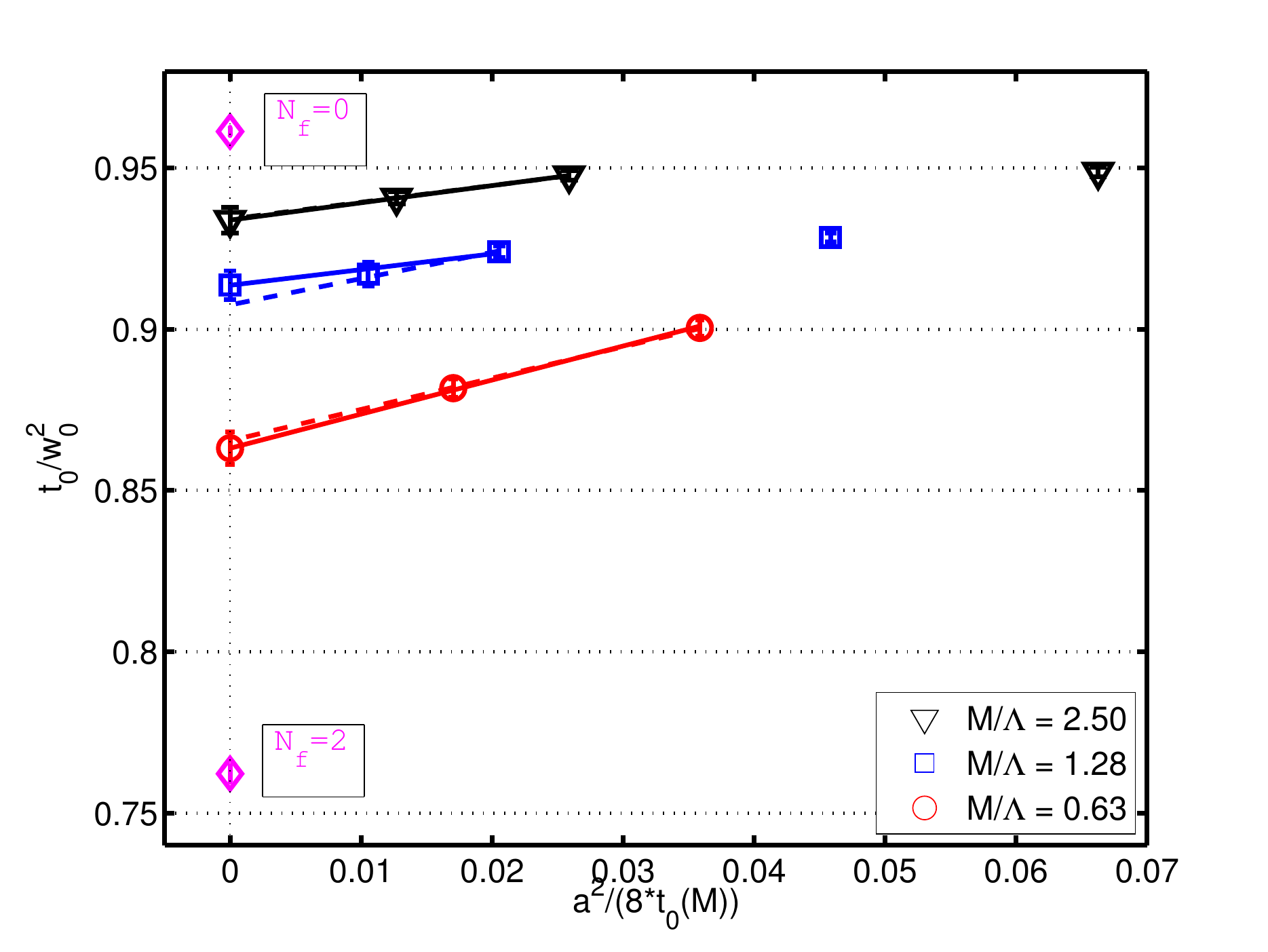}} \ \
  \resizebox{7cm}{!}{\includegraphics{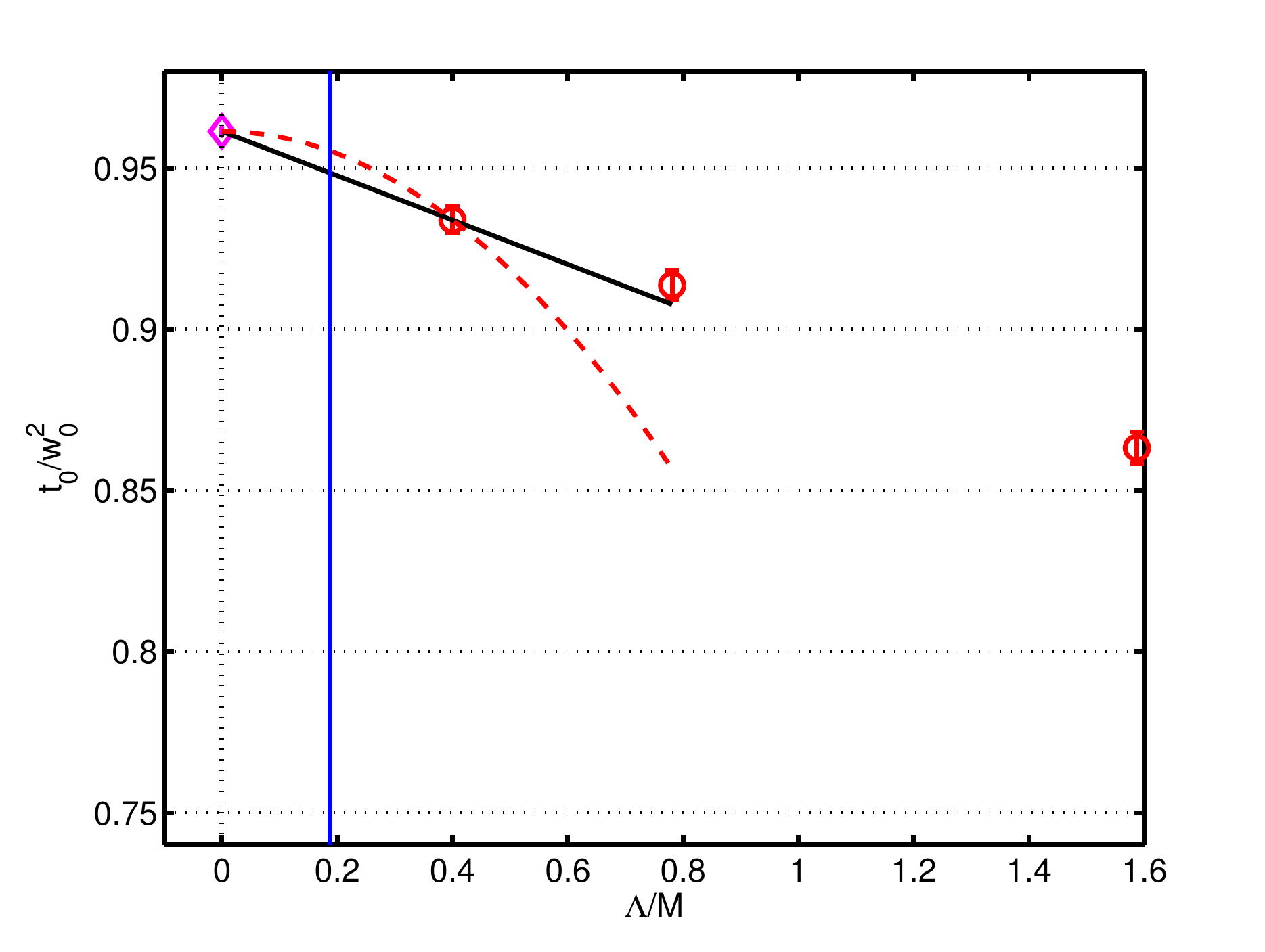}}
  \caption{Continuum extrapolation of the ratio $t_0/w_0^2$ (left) and
its mass dependence (right).}
  \label{f:t0ow0sq}
\end{figure}

In order to study the effects of heavy sea quarks,
we pick low energy gluonic observables with a strong dependence 
on the number $\Nf$ of sea quarks \cite{Bruno:2013gha}.
For example we consider the scales
$t_0$ \cite{flow:ML}
and $w_0$ \cite{flow:w0} defined from the action density ${\mathcal E}(t)=t^2\ev{E(x,t)}$, where $t$ is the Wilson flow time, through
\begin{eqnarray}
t_0 & : & {\mathcal E}(t_0) = 0.3 \,, \\
w_0 & : & t {\mathcal E}'(t)|_{t=w_0^2} = 0.3 \,.
\end{eqnarray}
Furthermore we take the scales $r_0$ \cite{pot:r0} 
and $r_1$ \cite{Bernard:2000gd}
defined from the static force $F(r)$ through
\begin{equation}
r^2F(r)|_{r=r_c} = c\,,\quad r_0\equiv r_{1.65} \,.
\end{equation}
Using these scales we form dimensionless ratios
\begin{equation}\label{e:ratios}
R=t_0/w_0^2\,,\; r_1/r_0\,,\;r_0^2/t_0 \,.
\end{equation}
We correct the values of the ratios for small differences
in the values of $M/\Lambda$, cf. \tab{t:ens}. The target values are
\begin{equation}\label{e:targetM}
\frac{M}{\Lambda} = 0.63\,,\; 1.28\,,\; 2.50
\end{equation}
and have been chosen to keep the size of the mass corrections small
at the finer lattice spacings.
The strategy is to fit the
data points on the finest lattices linearly in $M/\Lambda$ and
in $\Lambda/M$ and then take the slopes from the fits to correct 
the data at the other lattice spacings.
The statistical error of the corrected data is augmented
by the difference between the two fits.
We neglect in the corrections the error on $M/\Lambda$ since it mainly 
comes from $\Lambda$ and is therefore common to all points.

\begin{figure}\centering
  \resizebox{7cm}{!}{\includegraphics{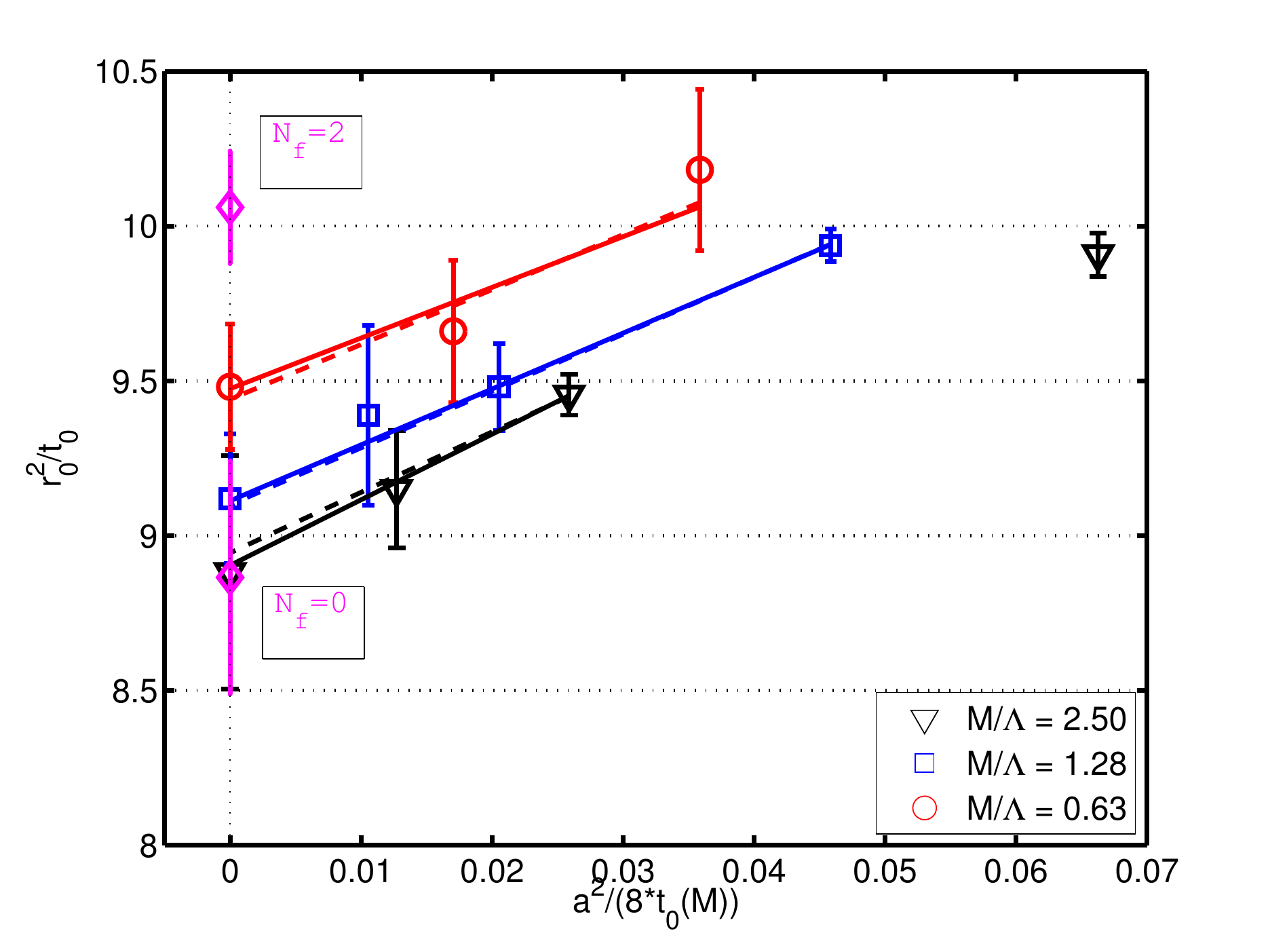}} \ \
  \resizebox{7cm}{!}{\includegraphics{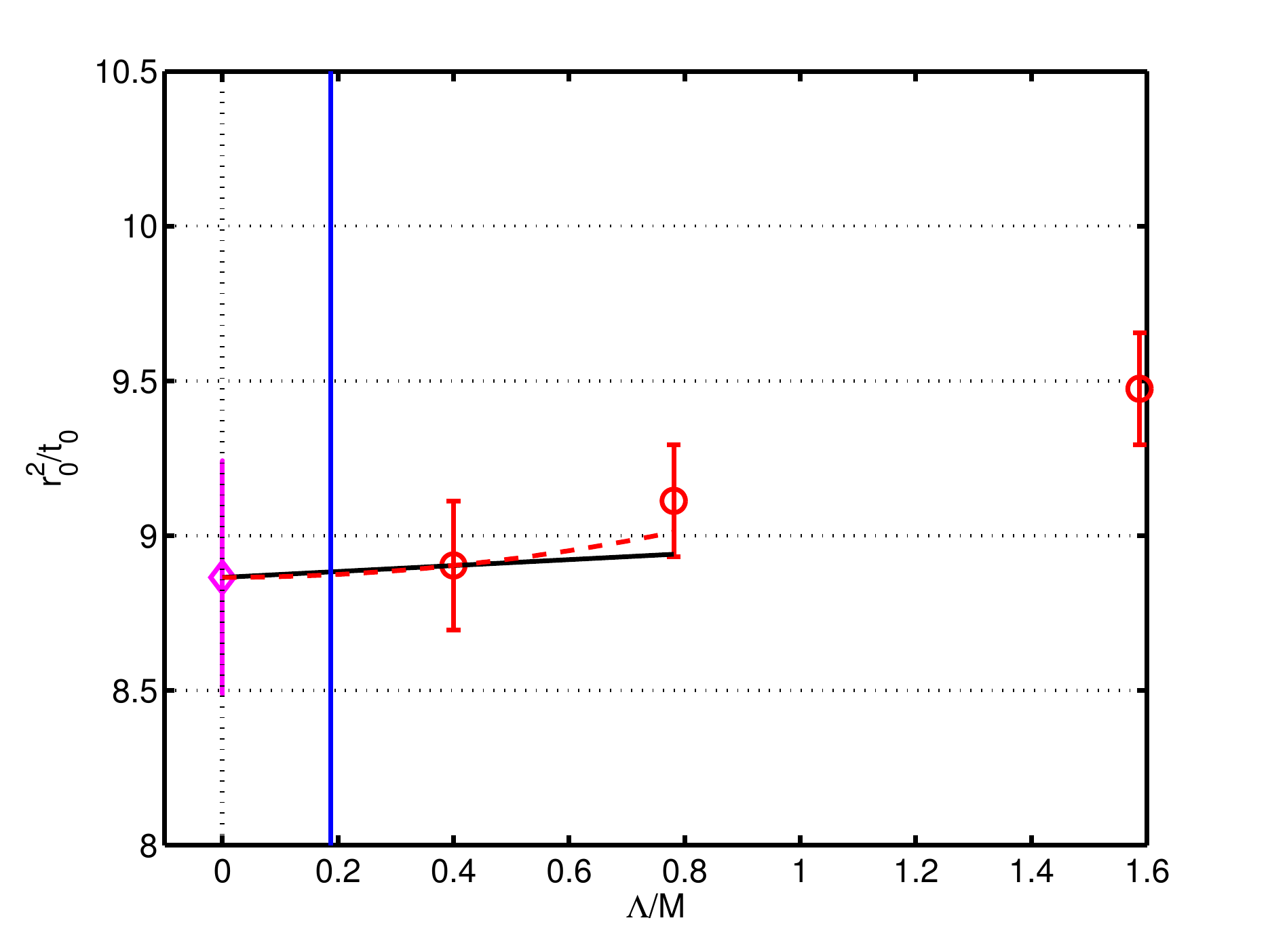}}
  \caption{Continuum extrapolation of the ratio $r_0^2/t_0$ (left) and
its mass dependence (right).}
  \label{f:r0sqot0}
\end{figure}

The continuum extrapolations are done by
global fits to the ratios $R_{\rm Lat}$ measured on the lattices,
see the left panel of \fig{f:t0ow0sq} for
$R=t_0/w_0^2$ and of \fig{f:r0sqot0} for $R=r_0^2/t_0$.
The continued lines correspond to fits where the slope $s$ of the 
$a^2$ effects at $M=0$ is fixed from \cite{Sommer:2014mea}
($s\approx2$ for $R=t_0/w_0^2$, $s\approx15$ for $R=r_0^2/t_0$)
\begin{equation}
R_{\rm Lat} = R(M) + s \frac{a^2}{8t_0}\left(1 + k_1 \frac{M}{\Lambda} + k_2 \frac{M^2}{\Lambda^2}\right) \,,
\end{equation}
where the continuum values $R(M)$ and $k_1$, $k_2$ are the fit parameters.
The dashed lines correspond to fits
\begin{equation}
R_{\rm Lat} = R(M) + k_0\frac{a^2}{8t_0} + k_1 \frac{a^2}{8t_0} \frac{M}{\Lambda}
\end{equation}
where the continuum values $R(M)$ and $k_0$, $k_1$ are the fit parameters.
The continuum values for $\Nf=0$ ($M/\Lambda=\infty$) are taken
from \cite{flow:ML} and our own results,
while the values for $\Nf=2$ at $M=0$ are from \cite{Bruno:2013gha}.
\begin{table}[t]
 \centering
\begin{tabular}{ccccccc}
\toprule 
 &  \multicolumn{6}{c}{$M/\Lambda$} \\
\cmidrule(rl){2-7}
$R$ & \multicolumn{2}{c}{$\Mc/\Lambda$} & 2.50 & 1.28 & 0.63  & $0$\\
\cmidrule(rl){2-3}
 & \small{$1/M$-scaled} & \small{$1/M^2$-scaled} & & & & \\
\midrule
$\sqrt{t_0}/w_0$ & 0.34(5)\% & 
0.16(2)\% & 0.72(11)\% & 1.26(12)\% & 2.62(14)\% & 5.4\%\\
\midrule
$r_1/r_0$        & 0.45(13)\% &
0.21(6)\% &  1.0(3)\%  &  1.8(5)\%  &  2.6(6)\% & $\approx$4\%\\
\midrule
$r_0/\sqrt{t_0}$    & 0.05(28)\% & 
0.02(12)\% &  0.1(6)\%  &  0.7(5)\% & 1.7(5)\% & 3\% \\
\bottomrule
\end{tabular}
\caption{Relative effects from decoupling of one heavy sea quark in ratios of
quantities of dimension one.
At $\Mc$ we quote the results from interpolations in $1/M$ 
and $1/M^2$.}
\label{t:releff}
\end{table}

The continuum extrapolated values are plotted as function of $\Lambda/M$
in the right plots of \fig{f:t0ow0sq} and \fig{f:r0sqot0}.
In order to estimate the size of the mass effects at the charm mass $\Mc$
(marked by a blue vertical line in the plots), we consider linear
(black continued lines) and quadratic (dashed red lines) interpolation
between our largest mass $M/\Lambda=2.50$ and the $\Nf=0$ ($M/\Lambda=\infty$)
values. The relative effects
\begin{equation}
 \label{e:releff} 
    {1 \over \Nf} {R(M)  - R(\infty) \over R(\infty)}
\end{equation}
are listed in \tab{t:releff} for ratios like
in \eq{e:ratios} but taken between quantities of dimension one.
The factor $1/\Nf$ is used to rescale the effects to the case of the 
decoupling of a single heavy quark $1\to0$.

\section{Conclusions}

Our numbers provide an estimate for charm effects in low energy observables
in $2+1+1$ simulations. 
As can be seen from \tab{t:releff} these effects are
very small, between 1 and 6 permille, in our model
of decoupling $2\to0$ heavy quarks (the numbers in \tab{t:releff}
are rescaled for decoupling $1\to0$).
This suggests that
tiny effects are being missed in $2+1$ simulations 
at low energies, given that
no qualitative difference between decoupling $2\to0$
and decoupling $2+1+1 \to 2+1$ is expected.
In this work we investigated low energies up to $r_1^{-1}$.\\

{\bf Acknowledgement.}
We are grateful for
computer time allocated for our project
on the Konrad and Gottfried computers at 
the North-German Supercomputing Alliance HLRN,
the Cheops computer at the University of Cologne (financed
by the Deutsche Forschungsgemeinschaft),
the Stromboli cluster at the University of Wuppertal
and the PAX cluster at DESY, Zeuthen.
This work is supported by the Deutsche Forschungsgemeinschaft
in the SFB/TR~09 and the SFB/TR~55.

\end{document}